\documentstyle[preprint,12pt,aps,psfig]{revtex}

\tighten
\begin{document}
\draft
\preprint{Imperial/TP/96-97/71}

\newcommand{\nc}{\newcommand}
\nc{\al}{\alpha}
\nc{\ga}{\gamma}
\nc{\de}{\delta}
\nc{\ep}{\epsilon}
\nc{\ze}{\zeta}
\nc{\et}{\eta}
\renewcommand{\th}{\theta}
\nc{\ka}{\kappa}
\nc{\la}{\lambda}
\nc{\rh}{\rho}
\nc{\si}{\sigma}
\nc{\ta}{\tau}
\nc{\up}{\upsilon}
\nc{\ph}{\phi}
\nc{\ch}{\chi}
\nc{\ps}{\psi}
\nc{\om}{\omega}
\nc{\Ga}{\Gamma}
\nc{\De}{\Delta}
\nc{\La}{\Lambda}
\nc{\Si}{\Sigma}
\nc{\Up}{\Upsilon}
\nc{\Ph}{\Phi}
\nc{\Ps}{\Psi}
\nc{\Om}{\Omega}
\nc{\ptl}{\partial}
\nc{\del}{\nabla}
\nc{\be}{\begin{eqnarray}}
\nc{\ee}{\end{eqnarray}}
\nc{\lambar}{\overline{\lambda}}
\nc{\plaq}{\Box}
\nc{\dumb}{\bullet \!\!-\!\!\bullet}
\nc{\plaqd}{\Box\!\!\!\!\Box_2}
\nc{\plaqc}{\Box\!\!\Box}
\nc{\dumbd}{\bullet \!\!-\!\!\bullet_2}
\nc{\dumbc}{\bullet \!\!-\!\!\bullet \!\!-\!\!\bullet}
\nc{\pldumb}{\Box \bullet \!\!-\!\!\bullet}
\nc{\dumbcirc}{\bullet \!\!-\!\!\bullet \circ}
\nc{\circd}{\circ \;\circ}

\title{On the Phase Structure of the $3D$ $SU(2)$--Higgs Model and the
       Electroweak Phase Transition}
\author{T. S. Evans\footnote{email: t.evans@ic.ac.uk},
        H. F. Jones\footnote{email: h.f.jones@ic.ac.uk}, 
        and A. Ritz\footnote{email: a.ritz@ic.ac.uk}\\ $\;$\\}
\address{Theoretical Physics Group, Blackett laboratory, \\
         Imperial College, Prince Consort Rd., 
         London, SW7 2BZ, UK.\\$\;$\\}
\date{\today}

\maketitle

\begin{abstract}
The phase structure of the 3D $SU(2)$--Higgs model, the dimensionally reduced 
effective theory for the electroweak model at finite temperature, is analysed
on the lattice using a variant of the linear $\de$--expansion. 
We develop a systematic 
variational cumulant expansion for general application to 
the study of gauge invariant operators in $3D$ gauge-Higgs
models, with emphasis on the symmetric phase. In particular, the technique
is not restricted to finite lattice volumes, and
application to the fundamental $3D$ $SU(2)$--Higgs model allows 
the discontinuity of certain observables across the first--order
transition to be observed directly for small $4D$ Higgs masses. The
resulting phase structure agrees well with Monte Carlo simulations
for small Higgs masses, but, at least to the order calculated, the technique 
is less sensitive to the expected evolution of the transition to a crossover 
for Higgs masses above 80 GeV.  
\end{abstract}
        

\section{Introduction}
Recently, considerable effort has been invested in the study of the
phase structure of the electroweak model and various extensions at
finite temperature. In particular, the properties of the first--order
phase transition are of great cosmological significance, in order
to determine whether the generation of baryon asymmetry is viable
at such a transition. 

The study of gauge theories, such as the electroweak model, 
at finite temperature is plagued by the
fact that combined with the need for non-perturbative techniques
to study the confining symmetric phase, infra-red effects also lead
to a large effective expansion parameter in the
broken phase where perturbation theory appears naively
applicable. This is particularly true for moderately large Higgs masses,
$m_H \sim m_W$. 
A considerable advance for the study of static, thermodynamic,
properties has been the systematic development of
dimensional reduction techniques \cite{kajantie96a}
allowing the mapping, via matching
of Green functions, of the full $4D$ theory onto a super-renormalizable
$3D$ theory corresponding to the Matsubara zero-modes. For the
electroweak theory in the transition region this procedure may
be carried out perturbatively as the gauge coupling is small, while
infra-red problems are also absent as such effects are encoded in
the dynamics of the effective $3D$ theory. If one ignores the 
unimportant $U(1)$ factor \cite{kajantie96d}, the effective theory
is a super-renormalizable $SU(2)$-Higgs model. This super-renormalizability, 
and also the non-triviality of the Higgs sector in $3D$, allow a more
straightforward approach to the continuum limit than in conventional
$4D$ studies \cite{evertz87,damgaard87,damgaard88,bock90,csikor96,aoki97}. 
Removal of
thermally massive fermions, and smaller $3D$ lattices have allowed detailed
Monte Carlo studies of the $3D$ theory, the conclusion being that
the first--order transition which grows increasingly weaker as the
Higgs mass increases, actually ceases to be first order and is presumably
an analytic crossover for 
$m_H\geq 80 $ GeV \cite{kajantie96c,karsch97,gurtler97}. This is plausible 
as in this system there is no gauge invariant order parameter which
can physically distinguish the two phases \cite{fradkin79}. Indeed recent
analysis of the spectrum for large Higgs masses \cite{philipsen96,philipsen97}
has indicated that the
structure is very similar on both sides of the transition/crossover region.
One can intuitively map three massive vector bosons and one scalar
of the Higgs phase smoothly onto the three vector and one scalar
bound state resonances of the confining phase. However, 
actual identification of the spectrum requires more care due to operator
mixing and the presence of low lying excited states.

Notwithstanding the success of Monte Carlo simulations, it is also
clear that many of the advantages of dimensional reduction for the
study of finite temperature gauge theories are more generally applicable.
Indeed, renormalization group \cite{reuter93,tetradis97}, and 
Schwinger-Dyson \cite{buchmuller94}, techniques
were among the first to suggest loss of the first--order transition,
stimulating much of the recent lattice activity. Furthermore, analysis
of the spectrum, and properties of the crossover very close to 
the endpoint of the first--order transition line, may well prove difficult
to study via Monte Carlo techniques due to the requirement for large
volume lattices in order to overcome finite size effects. This is 
especially true if the transition line ends at a second--order endpoint.

In this paper we shall investigate the fundamental 
$3D$ $SU(2)$--Higgs model on the lattice using an analytic technique 
which is a variant of the linear $\de$--expansion (LDE). The approach is
developed as a systematic method applicable to the calculation
of gauge invariant expectation values in gauge--Higgs theories
on lattices of arbitrary (even infinite) volume. We concentrate
in this instance on the fundamental $SU(2)$--Higgs model as 
the dimensionally reduced effective theory for the electroweak model
and consider the calculation of average plaquette and 
hopping term  expectation values in order to extract the phase structure.
Some preliminary results of this work were presented in \cite{ejr97a}.
A similar methodology has been used previously 
for studying the average plaquette energy in $4D$ lattice gauge 
theory \cite{zheng87,tan89,duncan88,duncan89,buckley92a,buckley92b,akeyo93a},
the phase structure in a mixed fundamental/adjoint $SU(2)$
model \cite{akeyo93b}, Higgs models \cite{zheng91,yang91,yang92}, and 
calculation of the scalar glueball mass in pure $SU(2)$ gauge 
theory \cite{akeyo95}. The results have generally agreed well with
Monte Carlo data where available.

The layout of the paper is as follows.
In section 2 we briefly review the relevant relations between 
the dimensionally reduced $3D$ theory and the 
electroweak $4D$ parameters, which allow us to present the
results in terms of $4D$ Higgs mass and temperature variables. 
In Section 3 we review the LDE approach, and describe the
application to gauge-Higgs systems, and the fundamental $SU(2)$-Higgs
model in particular. The calculational techniques are discussed
in section 4, with more technical details relegated to two appendices.
Results for relevant observables in the
transition region are presented, with higher order cumulants being used to 
help locate the critical line precisely; the results agree remarkably
well with Monte Carlo estimates. Some concluding remarks are presented
in Section 5.

\section{Dimensional reduction of the Electroweak Model at Finite temperature}
A full analysis of dimensional reduction of various $4D$ theories including
the standard model and MSSM, involving detailed matching of Green
functions has been presented in \cite{kajantie96a}. Here we shall simply
summarise the procedure and present the relations between the
parameters for later use.

The reduction procedure consists of two stages. In the first, high momentum
modes of the gauge field, and the thermally massive fermions 
are integrated out, while retaining the temporal component of the 
gauge field $A_0$ as an additional
adjoint Higgs. The mass of this field is given by the Debye screening mass
as $O(gT)$, and this field may also be integrated out as a second step. One then
matches the resulting Green functions with those for a 
super-renormalizable $3D$ theory which, neglecting the $U(1)$ factor,
is an $SU(2)$-Higgs theory,
\be
 {\cal L} & = & \frac{1}{4}G_{ij}^aG_{ij}^a + (D_i\Ph)^{\dagger}
      (D_i\Ph) + m_3^2\Ph^{\dagger}\Ph + \la_3(\Ph^{\dagger}\Ph)^2,
   \label{contin}
\ee
where $G_{ij}^a$ is the $SU(2)$ field strength, and $\Ph$ an $SU(2)$
doublet. More recently, study of the full $SU(2)\times U(1)$ theory
has indicated \cite{kajantie96d} that the phase structure is not 
qualitatively changed, thus for convenience we consider only the
case where the Weinberg angle is set to zero. It is an important feature
of the
Green function matching procedure that it does not introduce additional 
non-local terms which generically arise in a low energy effective theory
obtained by explicitly integrating out massive modes,
and may give dominant contributions. It has been shown \cite{kajantie96a}
that the Green functions of (\ref{contin}) approximate
those for the full theory in the infra-red up to corrections
of the form $\de G/G\sim O(g^3)$ where $g$ is the $4D$ gauge coupling. 
This is an acceptable approximation as the phase transition region 
is expected to have a weakly coupled gauge sector.

The $3D$ gauge coupling $g_3^2$, Higgs mass $m_3^2$ and self coupling
$\la_3$ will depend on the temperature and the underlying
$4D$ parameters. In this theory all parameters are dimensionful
and one may fix the scale with the gauge coupling $g_3^2$. The
phase structure and $4D$ temperature dependence are then determined by the
two dimensionless ratios \cite{kajantie96a}
$x \equiv \la_3/g_3^2$ and $y \equiv m_3^2(g_3^2)/g_3^4$. However, the
$3D$ theory is an effective theory for a large class of $4D$ theories
with differing scalar and fermionic field content, and the mapping
needs to be established for each individually. As discussed above, this may be 
performed perturbatively as calculations only involve massive modes,
and are thus free of infra-red problems. For the standard model the
relevant mapping has been determined in \cite{kajantie96a}. For simplicity,
we consider the mapping to the effective $4D$ parameters $m_H^*$ and $T^*$
of the $4D$
$SU(2)$-Higgs model \cite{kajantie96b,kajantie96c} which are convenient
for representing the results, and differ by only a few percent from 
the pole mass and temperature. 
Using tree--level relations \cite{kajantie96b}, the temperature
is given purely by the gauge coupling, $g_3^2 = 0.44015 T^*$, while
the dimensionless ratios are
\be
 x & = & -0.00550+0.12622 h^2 \\ \label{x}
 y & = & 0.39818+0.15545 h^2-0.00190 h^4 - 2.58088\frac{(M_H^*)^2}{(T^*)^2},
               \label{y}
\ee
where $h=m_H^*/80.6 $ GeV is a dimensionless Higgs mass. As 
there is this direct
connection between $(m_H^*,T^*)$ and $g_3^2,x$, and $y$ we can convert analysis
of the phase structure in the $x,y$ plane to the physically more 
meaningful $(m_H^*,T^*)$ variables.

Focusing now on the $3D$ theory, the fact that the infrared details
are still encoded in the dynamics of this theory means that a 
non-perturbative approach is required. The super-renormalizability
of (\ref{contin}) now ensures that one can calculate perturbatively
exact RG trajectories. Only the mass parameter receives contributions 
at two-loop order, and there are no three-- or higher--loop contributions.
We adopt a lattice regularization, and the standard Wilson
lattice analogue of the action 
(\ref{contin}) with lattice constant $a$
may be written in the form
\be
 S & = & \beta\sum_p \frac{1}{2}{\rm Tr}U_p + \frac{1}{2} \beta_h \sum_{l_{ij}}
        \rh_i\rh_j {\rm Tr}U_{ij}
         -\sum_i \left[\rh_i^2+\beta_r(\rh_i^2-1)^2\right], \label{latt}
\ee
where we choose to represent the Higgs doublet in the form
$\Ph = \rh_i V_i$, with
$\rh\in${\bf R}$_+$ and $V$ an element of the fundamental 
representation of $SU(2)$. We then make a gauge transformation 
$U_{ij}\rightarrow V_iU_{ij}V_j^{\dagger}$ to rotate the phase of the 
Higgs field in the interaction term to the identity at each site. Thus the
interaction term becomes $\rh_i\rh_j$Tr$U_{ij}$. While this
choice may not appear desirable in the symmetric phase, it is
advantageous for calculational purposes, and in previous lattice studies
\cite{zheng91}
has led to good results, so we shall also adopt it here.
In the continuum limit, which may be determined by the RG trajectory
noted above, the dimensionless
parameters $\beta,\beta_h,\beta_r$ are related to the parameters of the
continuum theory as follows \cite{kajantie96b}:
\be
 x & = & \frac{\beta \la}{\beta_h^2}\;\;\;\;\;\;\;\;\;\;\;\;\;\;\;\;\; 
    g_3^2 a = -\frac{4}{\beta} \label{xlat} \\ 
 y & = & \frac{\beta^2}{8}\left(\frac{1}{\beta_h}-3
      -\frac{2\beta_r}{\beta_h}\right)
    +\frac{3\Si\beta}{32\pi}(1+4x) \nonumber\\
   &  & +\frac{1}{16\pi^2}\left[
    \left(\frac{51}{16}+9x-12x^2\right)\left(\ln\frac{3\beta}{2}+\zeta\right)
    +5.0+5.2x\right]  \label{ylat},
\ee 
where $\Si=3.17591$ and $\zeta=0.09$ follow from perturbative analyses. 
For different
values of the lattice constant $a$ these relations define a constant
physics curve, or RG trajectory, for fixed $g_3^2,x,y$ in the 
space $\beta,\beta_h,\beta_r$.

With the preceding results, we may now analyse the phase structure
of the lattice $SU(2)$--Higgs model in $3D$ and subsequently
convert the results
back to physical parameters. The details of this procedure
will be discussed in Section 4, while in the next section we review
details of the LDE, and discuss its application to the
$SU(2)$-Higgs model.

\section{LDE analysis of the $SU(2)$--Higgs Model}
The linear $\de$--expansion is an analytic approach to calculations
in field theory which  makes use of an artificial expansion not
dependent on the existence of small coupling constant for its validity.
Nonetheless, calculations are generally 
no more complex than standard perturbative
Feynman diagrams. An essential feature of this approach is the ability
to optimize convergence of the series by fixing additional parameters
appearing in the extended action.

The LDE action has the form
\be
 S_{\de} & = & (1-\de)S_0(J) + \de S,
\ee
where $S_0$ contains some dependence on a variational parameter $J$, and 
$S_{\de}\rightarrow S$ the theory under consideration, independent of $J$, 
when $\de=1$. The generating
functional for Green functions may then be expanded to an appropriate
order in $\de$, which is then set to unity.

As the power series is only calculated to a finite order, it retains
some dependence on $J$ which would be absent in a full summation. A well
motivated criterion for fixing $J$ is the principle of minimal
sensitivity (PMS) \cite{stevenson81}, whereby $J$ is chosen at a local
extremum of a physical quantity. i.e. if $X_N(J)$ denotes the $N$'th
approximation to $X$, then we impose
\be
 \frac{\ptl X_N(J)}{\ptl J} & = & 0.
\ee

This or a similar criterion is intrinsic to the success of the LDE,
providing the nonperturbative dependence on the coupling constant,
and has been shown to induce convergence of the series in 0- and
1-D field theories \cite{buckley93,duncan93} where the perturbative series
are asymptotic and eventually diverge factorially.

The technique has also been applied with success to 
lattice gauge theories,
where different choices of $S_0$ are appropriate in the weak and strong
coupling regimes. In the strong coupling regime an approach using a 
maximal tree of plaquettes \cite{duncan88,buckley92b} has proved successful
in describing the strong coupling behaviour of the average plaquette energy
$E_P$. The weak coupling regime has also been considered using
a quadratic $S_0$ \cite{duncan89,buckley92a} and using a single
link $S_0$ in \cite{zheng87}. These
techniques have since been applied at finite temperature \cite{tan89},
to the $SU(2)$ mass gap \cite{akeyo95}, mixed $SU(2)-SO(3)$ phase structure
\cite{akeyo93b}, and also to lattice Higgs models 
in \cite{zheng91,yang91,yang92}.

In the present context, the variational cumulant expansion approach
using a single link $S_0$ seems most appropriate. At this stage we
retain the full gauge Higgs coupling and choose a background
action of the form (cf. (\ref{latt}))
\be
 S_0 & = & \sum_{l} (J+\beta_h\rh_i\rh_j)\frac{1}{2} {\rm Tr} U_l -
    \sum_i \left[\rh_i^2+\beta_r(\rh_i^2-1)^2\right],
\ee
thereby introducing a variational parameter $J$. It should
be noted that this background
is not explicitly gauge invariant due to the single link action. 
In the present context this is not of concern since we shall
only calculate the expectation value of gauge invariant operators.
However, if one were to calculate
gauge non-invariant expectations the correct approach \cite{tan89}
is to note that a priori $J$ is link dependent and may have an
arbitrary $SU(2)$ phase which should be summed over. Once this procedure is
carried out one finds that gauge non-invariant expectation values vanish as
expected\footnote{It is interesting to 
note that this approach does in principle
allow the possibility of considering gauge fixed operators, and this may
be of interest in comparing gauge invariant lattice results for the
spectrum with gauge fixed perturbative calculations 
\cite{frohlich81,karsch96,buchmuller97}.}.
Using the methodology just described we evaluate expectation values
as follows:
\be
 \left<X\right> & = & \frac{1}{Z}\int [dU][d\rh] X e^{\de(S-S_0)}e^{S_0},
\ee
where the partition function is given by
\be
 Z = e^{-W} &=& \int [dU][d\rh] e^{\de(S-S_0)}e^{S_0},
\ee
in which $[dU]$ denotes the standard Haar measure over $SU(2)$, and 
$[d\rh]=\prod_i {\mbox d}\rh_i \rh_i^3$.
An $N^{th}$ order approximant to $\left<X\right>$ in the LDE
then takes the form,
\be
 \left<X\right>_N & = & \sum_{n=0}^N \frac{\de^n}{n!}
     \left<X(S-S_0)^n\right>_C,
\ee
where $\left<X\right>_C$ denotes the connected expectation value,
or cumulant, of $X$ in the $S_0$ background. e.g. 
$\left<AB\right>_C=\left<AB\right>_0-\left<A\right>_0 \left<B\right>_0$.
It is the expansion of the partition function in the denominator
of the expectation value to the appropriate order in $\de$ that
naturally subtracts the disconnected components. This is
crucial, as one now observes that the physical expectation value
reduces to a sum of connected expectation values for which one can
introduce a convenient diagrammatic notation in order to keep track of
the independent contributions. Furthermore, connectedness implies that one
can effectively take the lattice volume to infinity as at any finite
order only a finite number of distinct diagrams will contribute.
Summing all such diagrams over an infinite lattice will then simply lead
to coefficients proportional to the number of sites $N$ in the lattice,
and thus expectation values normalised as $\left<X\right>/N$ will
be finite and calculable, independent of the finiteness or
otherwise of $N$.

Therefore, for practical purposes the calculation reduces to the
enumeration and calculation of expectation values of
connected diagrams in the $S_0$ background, i.e. 
we require\footnote{Note that if convenient one may reduce the 
number of diagrams required by calculating $\left<X(S^n)\right>_0$
in an $S_0(1-\de)$ background, and then subsequently expand the
result to the appropriate order in $\de$.}
\be
 <X>_0 & = & \frac{1}{Z_0}\int [dU][d\rh] X e^{S_0},
\ee
with
\be
 Z_0 & = & \int [dU][d\rh] e^{S_0},
\ee

Actual calculations are simplified by recognising that the terms in the
$SU(2)$--Higgs action involve only real constants, the $\rh_i$, or
$SU(2)$ characters since Tr$U=\ch_{1/2}(U)$ for $U$ an element of 
the fundamental 
representation of $SU(2)$. The free action $S_0$ also consists
only of constant factors and fundamental $SU(2)$ characters. For example,
the $S_0$ partition function takes the form,
\be
 Z_0 & = & \int [\prod_i {\mbox d}\rh_i \rh_i^3]\int [\prod_{l_{ij}}{\mbox d}
           U_{ij} \exp\left(\sum_l(J+\beta_h\rh_i\rh_j)\ch_{1/2}(U_{ij})
               -\sum_i V_H(\beta_r,\rh_i)\right),
\ee
where $V_H=\rh^2+\beta_r(\rh^2-1)^2$ is the Higgs potential.
The expectation value integrals are most easily evaluated making use of 
a character expansion for the exponentiated $S_0$. For the fundamental
representation of $SU(2)$ we have (see e.g. \cite{halliday84})
\be
 e^{\frac{1}{2}J\ch_{1/2}(U)} & = & \frac{2}{J}\sum_{r=0}(r+1)I_{r+1}(J)
         \ch_{r/2}(U),
\ee
where we have made use of the relation $I_p(x)-I_{p+2}(x)=2(p+1)I_{p+1}(x)/x$
for modified Bessel functions. Then via use of the orthogonality relation
for characters the group integration becomes trivial. In certain
higher order cases expectation values do not reduce to simple
products of link characters and a rather more general technique 
is then required
(see Appendix A for more details). In order to make the calculations tractable
we will make the assumption that 
the Higgs modulus varies relatively slowly over the lattice. More
precisely, we assume that $\rh_i\rh_j=\rh_i^2+\de^2\rh_i\De\rh_i$, where
$\De\rh_i=\rh_j-\rh_i$ with $i$ and $j$ nearest neighbour sites.
This is motivated by the success of the approach, ignoring 
the correction altogether,
in spin-glass systems \cite{damgaard85}, which are very similar
to the $\beta\rightarrow 0$ limit of the $SU(2)$-Higgs lattice theory.
The validity of this assumption for large $\beta$ is less clear. However,
given our ansatz, at higher order in $\de$ the correction terms enter in the 
calculation in a self--consistent way. Consequently, up to 
$O(\de^2)$, once the group integration has been performed via character
expansion techniques, the remaining integrals over
the modulus of the Higgs field generically 
take the form (for each link)
\be
 A^n_r & \equiv & \left[\int_0^{\infty}{\mbox d}\rh \rh^{2n+3}e^{-V_H}
              \left((r+1)\frac{I_{r+1}(J+\beta_h\rh^2)}{\frac{1}{2}
               (J+\beta_h\rh^2)}\right)^d\right]^{1/d},
         \label{Anr}
\ee
which may be calculated numerically (see Appendix B for an analytical study, 
and an expansion in $\beta_h$). All the 
required expectation values may then be expressed in terms of the 
ratios $B^n_r\equiv A^n_r/A^0_0$.

\section{Expectation Value Calculations and the Phase Structure}
In considering expectation values it is convenient to introduce
a graphical notation for the standard operators appearing in the
action. We define:
\be
 \plaq & = & \sum_p\frac{1}{2}{\rm Tr} U_p \\
 \dumb & = & \sum_l\frac{1}{2}\rh_i\rh_j {\rm Tr} U_{ij} \\
  -\!\!\!- & = & \sum_l\frac{1}{2}{\rm Tr} U_l.
\ee
The gauge invariant quasi-order parameters,
the average plaquette, and the hopping term, may then be
conveniently represented in the form 
\be
 \left<E_P\right> & = & \frac{1}{N_p}\left<\plaq\right> \\
 \left<E_L\right> & = & \frac{1}{N_l}\left<\dumb\right>,
\ee
where the expectation value in this case corresponds to the full action
and, working with an $N$--site $d$--dimensional 
lattice, the number of links and plaquettes
is given by $N_l=Nd$, and $N_p=d(d-1)N/2$, respectively.

Disconnected contributions to these observables in the $S_0$ background
are calculated in the manner introduced in Section 3, and elaborated 
further in Appendix A. 
The required expectation values, along with the associated
multiplicities, of the relevant diagrams are also tabulated in 
Appendix A for reference.
In order to present reasonably concise expressions we refer to the
relevant connected contribution from 
diagram $di$ as $C_i=m_i\left<D_i\right>_C$
where $m_i$ is the multipicity and $\left<D_i\right>_C$ is the 
connected expectation value obtained by subtracting the disconnected
components from $\left<D_i\right>_0$ in the standard manner. 

We consider first the pure gauge sector of the theory. It is clear
that in the limit $\beta_h\rightarrow 0$ the partition function
reduces to that of pure $SU(2)$ gauge theory multiplied by an
overall factor determined by the integration over the Higgs modulus,
which factors out of expectation values. We now consider the
average plaquette for non-zero $\beta_h$ in the symmetric phase. To
first order in $\de$ the expectation value is given by
\be
 \left<E_P\right> & = & \frac{1}{N_p}(\left<\plaq\right>_0 
           +\de\left<\plaq(S-S_0)\right>_C+O(\de^2)).
\ee
Recalling that only connected diagrams contribute, the contribution
to $O(\de)$ is given by
\be
 \left<E_P\right>_1 & = & C_1+\beta (C_6 + C_7)-J C_8. 
\ee

The variational parameter must now be fixed by using the 
PMS criterion described earlier. Considering for a moment
a general observable $\left<{\cal O}(\beta,\beta_h,\beta_r,J)\right>$,
the procedure used throughout is to examine $\left<{\cal O}\right>$ at fixed
$\beta,\beta_h,\beta_r$ as a function
of $J$, fixing the parameter at the ``flattest''
local extremum $J=\tilde{J}$ in accordance with PMS. 
As a result, the expectation value
has the functional dependence $\left<{\cal O}\right>=
\left<{\cal O}(\beta,\beta_h,\beta_r,
\tilde{J}(\beta,\beta_h,\beta_r))\right>$ and, due to 

\begin{figure}
 \centerline{%
   \psfig{file=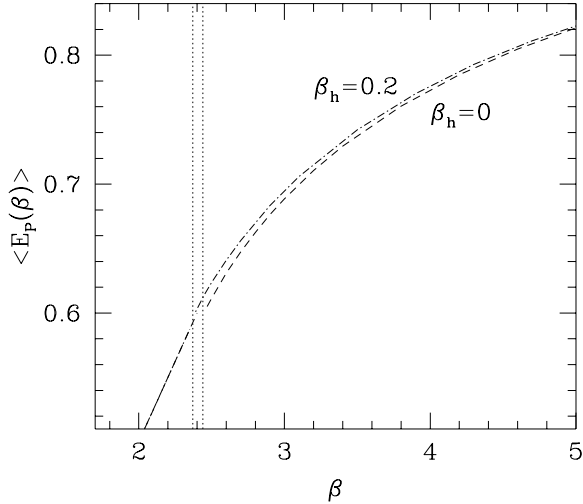,width=8cm,angle=0}%
   }
 \caption{The plaquette energy $E_p$ as a function of $\beta$ for
 ($\beta_h=0$, $\beta_r=0$) and also ($\beta_h=0.2$, $\beta_r=0.5$), a 
 smooth transition 
 between PMS branches occurring at $\beta\approx 2.44$ and 2.37
 respectively.} \label{P}
\end{figure} 

\noindent discontinuous
jumps between the PMS extrema $\tilde J$, may exhibit non-analyticities
even when $\left<{\cal O}(\beta,\beta_h,\beta_r,J)\right>$ is an 
analytic function of its parameters. Thus one associates the
non-analyticity of discontinuous observables across a first
order transition with jumps in the PMS solution $\tilde J$. Therefore
the appearance or disappearance of a particular $\tilde J$ as one moves
over the parameter space is associated with a phase transition of the
system if the resulting observable exhibits a discontinuity. This
interpretation is justified by the experience with this technique
in lattice gauge theory and in particular the study of the
$SU(2)-SO(3)$ phase diagram \cite{akeyo93b}, where the results agreed 
very well with the expectations from Monte Carlo simulations. 

Returning now to the average plaquette expectation, we plot the
results in Fig.~1 as a function of $\beta$ for $\beta_h=0$
(pure gauge theory) and $\beta_h=0.2$ (well into the symmetric phase).
One observes that the results are remarkably similar, indicating that
the presence of the Higgs sector  has little effect. 
This conclusion is consistent with explicit analysis of the 
$\left<E_P\right>$ expectation values (see Appendix B), which suggests that the
first Higgs sector corrections in the symmetric phase are generically
of $O((\beta_h/2)^4)$. 

The structure
of the free action $S_0$ indicates that the limit $J=0$ corresponds to the 
standard strong coupling expansion, and indeed a PMS solution exists
for all $\beta$ at $J=0$. A second PMS solution first appears
at $\beta\sim 2.4$, then flattens out to become the clear PMS point as
$\beta$ increases. This transition to a ``weak coupling'' branch
occurs, however, at almost precisely the same value of $\left<E_P\right>$,
and with no distinctive change in behaviour,
implying that the transition is actually a crossover. This deconfinement 
crossover is indeed consistent with expectations from lattice Monte Carlo
studies in $4D$ \cite{damgaard87,evertz87,bock90}, 
as is the slight shift of the crossover to lower
values of $\beta$ as $\beta_h$ increases.

We now turn now to consideration of the hopping term which, for small
Higgs masses, serves as a useful gauge invariant operator which
is discontinuous across the first--order transition. In order to analyse the
behaviour as precisely as possible we take the calculation 
to $O(\de^3)$,
\be
 \left<E_L\right> & = & \frac{1}{N_l}\left(
         \sum_{n=0}^3\frac{\de^n}{n!}\left<\dumb(S-S_0)^n\right>_C 
           +O(\de^4)\right).
\ee
The relevant connected diagrams and their expectation
values are enumerated in Appendix A. The result is given by
\be
 \left<E_L\right> & = & C_1 +\de\left(2(d-1)\beta C_4-J C_5\right) \nonumber\\
    & &  +\frac{\de^2}{2}\left(\beta^2\frac{d-1}{2}(C_{10}+C_{11}+C_{12})
                -\beta J(d-1)(C_{13}+C_{14})+J^2C_{15}\right) \nonumber\\
    & &  +\frac{\de^3}{6}\left(\beta^3\frac{d-1}{2}
               \sum_{i=27}^{36}C_i-3\beta^2 J\frac{d-1}{2}\sum_{i=37}^{43}C_i
                          \right.  \nonumber\\
    & & \;\;\;\;\;\;\;\;\;\;\;\;\;\;     
      \left.+3\beta J^2\frac{d-1}{2}(C_{44}+C_{45}+C_{46})-J^3C_{47}\right)
+O(\de^4).
    \label{el}
\ee
Considering first the expectation value at $O(\de)$ only, we indeed find
a transition between PMS branches in the region of $\beta_h\approx 0.31-0.38$
consistent with Monte Carlo expectations for the first--order transition.
However, there is a region in parameter space in the transition region
between these two branches for which no PMS extrema exist. This
is not unexpected, as the expectation value undergoes a discontinous
shift. However, it does mean that a precise determination of the
critical point will prove difficult working directly from $\left<E_L\right>$.
Analysing a range of $\beta$ values, the results are presented in Fig.~2
for a $4D$ Higgs mass of $70$ GeV, where this constraint determines $\beta_r$
via (\ref{x}) and (\ref{xlat}). Since the approximation
used in evaluating the integral over the Higgs modulus is expected to be most
accurate in the $\beta\rightarrow 0$ limit,
we restrict ourselves to relatively small values of $\beta$ in this
particular calculation.

\begin{figure}
 \centerline{%
   \psfig{file=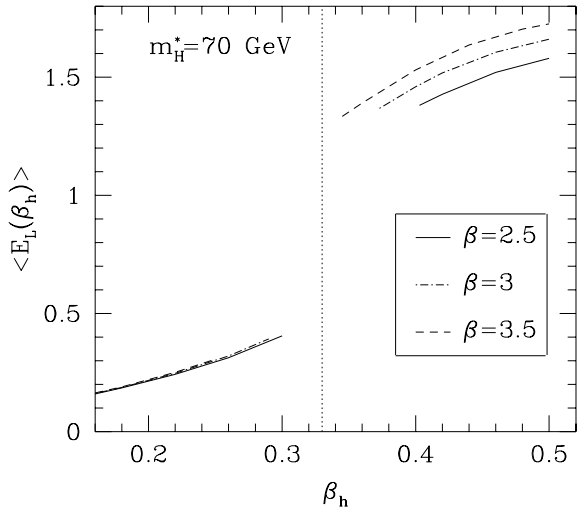,width=8cm,angle=0}%
   }
 \caption{We plot $<E_L>_1$ versus $\beta_h$ for a $4D$ Higgs mass
 $m_H^*=70$ GeV, and three values of $\beta$. The vertical line is simply
 a rough guide to the position of the transition region
 $\beta_h=0.31-0.35$.}\label{L}
\end{figure} 

The results presented in Fig.~2 are extended to $O(\de^3)$ in Fig.~3.
One observes that the transition region is now more clearly defined, with the
PMS branches persisting closer to the transition point itself.
At this order one has additional terms from diagrams
at $O(\de^0)$ ($d1$), and $O(\de^1)$ ($d4$ and $d5$), 
correcting the slowly varying Higgs approximation
in the evaluation of $A^n_r$.
However, we note that in the transition region the behaviour of
$\left<E_L\right>$ is dominated by long wavelength modes, and thus
the approximation used should be appropriate. 
A sample analytic calculation of the correction term is presented in
Appendix B, and provided one assumes that the modulus of the Higgs field
varies by no more than 10\% over a lattice spacing, one finds that
the correction terms are negligibly small when compared with the
original contribution at the appropriate order in $\de$.

\begin{figure}
 \centerline{%
   \psfig{file=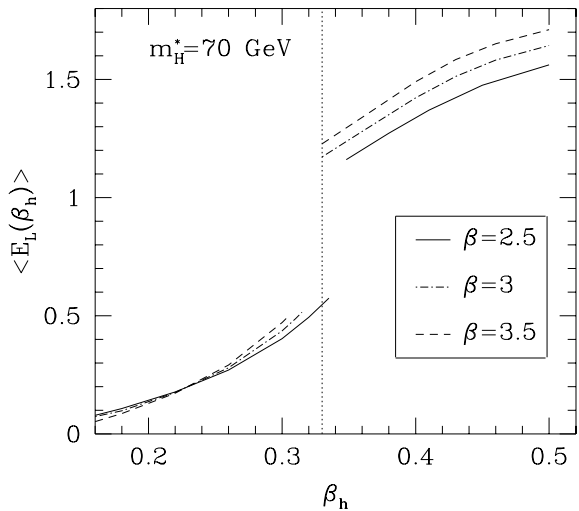,width=8cm,angle=0}%
   }
 \caption{We plot $\left<E_L\right>_3$ versus $\beta_h$ for a $4D$ Higgs mass
 $m_H^*=70$ GeV, and three values of $\beta$. The vertical line is at the
 same value of $\beta_h$ as in Fig.~2.}
 \label{L2}
\end{figure}

In order to determine the transition point more
precisely, we now consider the second--order cumulant of
$E_L$, defined by $C(E_L)\equiv \left<E_L^2\right>-\left<E_L\right>^2$.
To $O(\de)$ this expectation may be represented as,
\be
 C(E_L) & = & \frac{1}{N_l}\left<\dumb^2\right>_C \\
    & = &  C_{48}+\de\left(\beta\frac{d-1}{2}(C_{49}+C_{50}+C_{51})
           -J(C_{52}+C_{53})\right)+O(\de^2).
\ee
The cumulant exhibits a PMS solution for small $\beta_h$, but as
$\beta_h$ is increased this solution is lost at $\beta_{hc}$,
above which no PMS extrema exist. Accordingly, we associate this
point with the phase transition point. Tracking this line
for increasing $\beta$ we obtain an estimate for the transition line 
along the RG trajectories in the phase diagram for 
$4D$ Higgs masses of 60 and $70$ GeV. 
The results are shown in Fig.~4 and compared
with known Monte Carlo points in the diagram from the
simulations presented in \cite{kajantie96b}, \cite{gurtler96},
and Lee-Yang zeros analysis in \cite{gurtler97}. Our procedure,
even at this low order, produces close agreement with the Monte Carlo
results. 

\begin{figure}
 \centerline{%
   \psfig{file=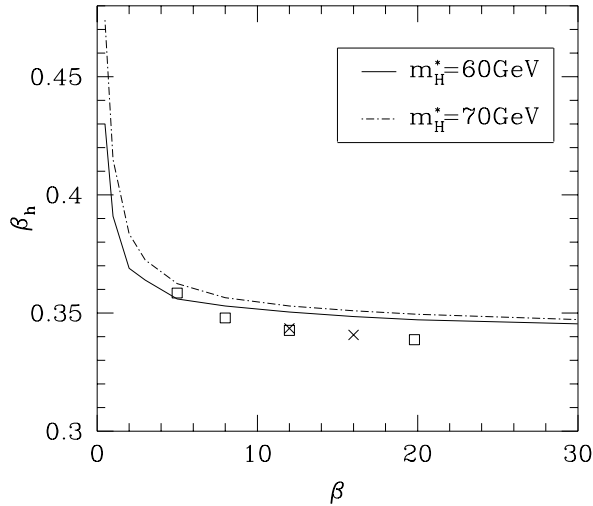,width=8cm,angle=0}%
   }
 \caption{We plot the critical curves in the $(\beta_h,\beta)$ phase diagram
 for Higgs masses of $60$ and $70 $ GeV. The critical values of $\beta_h$ were 
 determined to be those points at which a PMS extremum for the cumulant
 $C(E_L)_1$ was just lost; the values corresponding to a discontinuity
 of the cumulant. The results are compared with the Monte Carlo results
 of \protect\cite{kajantie96b} (squares), and \protect\cite{gurtler96}
 and \protect\cite{gurtler97} (crosses) for $m_H^*=60 $ GeV.}
 \label{ph}
\end{figure}

Extrapolating to the continuum limit along the RG trajectory,
we obtain critical couplings for each Higgs mass.
Using (\ref{y}) and (\ref{ylat}) we can also determine the
critical temperature and perform a similar extrapolation to the continuum.
However, as one observes from Fig.~4, the continuum limit of $\beta_{hc}$
is approximately 3\% above the Monte Carlo estimates. Although this
is entirely satisfactory for the order of the calculation, the sensitivity
of the critical temperature to the value of $\beta_{hc}$ 
results in estimates for the
critical temperature being significantly lower than one would expect.
In the low $\beta$ region, however, where the results at this order are 
expected to be most accurate, the critical temperature agrees with Monte Carlo
results \cite{kajantie96b} up to 20-25\%. Note also that while the 
first--order cumulant appears to over-estimate the critical coupling
$\beta_{hc}$, the transition region determined by analysis
of $\left<E_L\right>$ gives an under-estimate, for the
relevant $\beta$ range, of 6--8\%. This discrepency provides a
reasonable estimate of the systematic errors of these results.

Increasing the Higgs mass beyond the regime where the first--order
transition is expected to end, we find that the results change  
qualitatively only in the sharpness of the transition. 
In Fig.~5 the observable $\left<E_L\right>_3$ is plotted
at $\beta=3$ for a $4D$ Higgs mass of $120$ GeV. One observes that 
the PMS solutions do not persist as far into the
transition region in comparison with the results at $m_H^*=70 $ GeV, 
and that the discontinuity itself is slightly smaller in magnitude. 
Although it is not possible from this result
alone to characterise the transition as an analytic 
crossover rather than a first--order transition, it is 
certainly consistent with the former interpretation, 
which is suggested by previous studies.

\begin{figure}
 \centerline{%
   \psfig{file=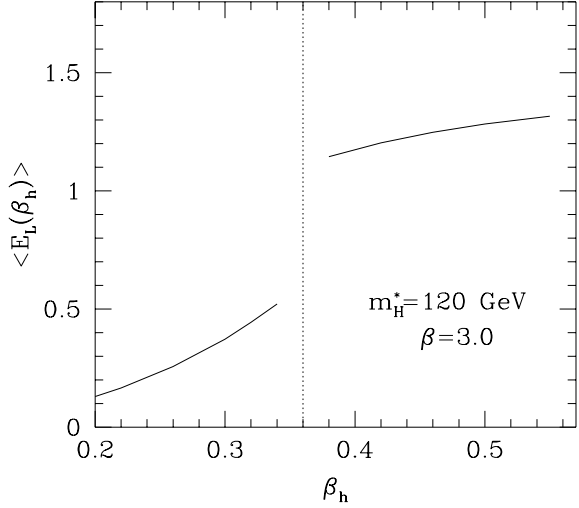,width=8cm,angle=0}%
   }
 \caption{We plot $<E_L>_3$ versus $\beta_h$ for a $4D$ Higgs mass
 $m_H^*=120$ GeV at $\beta=3$. The vertical line indicates
 the transition/crossover region.}
\end{figure}

\noindent The results are also consistent with the fact that
Monte Carlo simulations 
indicate that the crossover is still extremely sharp for a Higgs mass
in this region. Fig.~6, in which the calculation is repeated at 
$m_H^*=140$ GeV, shows a further reduction in the sharpness of
the transition, as one observes from the increase in the parameter
range for which no PMS solution exists.

\begin{figure}
 \centerline{%
   \psfig{file=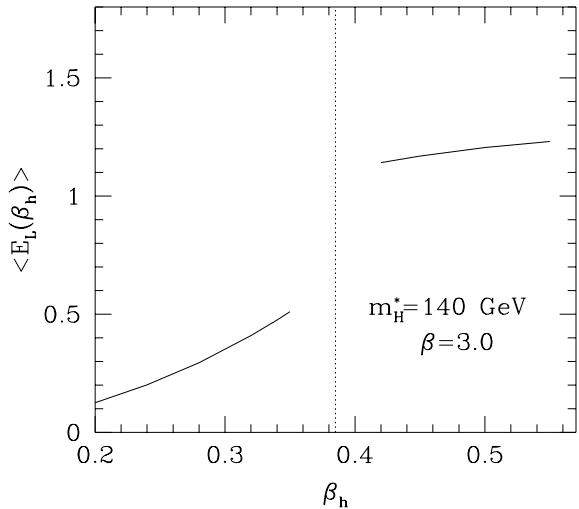,width=8cm,angle=0}%
   }
 \caption{We plot $<E_L>_3$ versus $\beta_h$ for a $4D$ Higgs mass
 $m_H^*=140$ GeV at $\beta=3$. The vertical line indicates
 the transition/crossover region.}
\end{figure}

Increasing the Higgs mass beyond $\sim 150 $ GeV leads to a loss
of the PMS solution in the broken phase. This is most likely to be
a finite order effect with little direct relevance for the true
phase structure, and for this reason
we shall ignore such large Higgs masses. In studying the
second--order cumulant, we find that the qualitative behaviour is again
unchanged until the Higgs mass becomes very large. In the region of interest,
$80-120 $ GeV, the loss of a PMS solution at a critical value of
$\beta_h$ may presumably be associated with the sharpness of the
crossover in this regime.

\section{Discussion}
In this paper we have made use of a variant of the linear $\de$--expansion
to examine the phase structure of the $SU(2)$-Higgs model in $3D$. 
Using the previously derived matching
conditions between the parameters of this model and a suitable
approximation of the full $4D$ finite temperature electroweak  theory
we were able to extract the phase structure,
consistent with lattice Monte Carlo simulations for small Higgs
masses. Increasing the Higgs mass beyond the point where
the first--order transition is expected to  be replaced
by a smooth crossover, the discontinuities in the observable
 $\left<E_L\right>$ persist, but become less pronounced, as $m_H^*$
increases. A higher order calculation in $\de$ may be necessary
in order to observe a more significant qualitative change in the
behaviour of the relevant observables.

Nevertheless, this work does suggest a number of immediate extensions 
which are currently under consideration. The formalism
developed here, in combination with the approach used in \cite{akeyo95},
allows an analytic study of the spectrum in the
symmetric phase, in particular near the critical endpoint, where
the lack of any finite volume restriction makes the regime dominated
by large correlation lengths quite accessible without detailed
finite volume scaling analysis. Recent work on the
spectrum of the $3D$ theory for large Higgs masses suggests
that in this respect the Higgs and confining phase have very similar 
properties \cite{philipsen97}. 

The technique also suggests the possibility of studying 
the adjoint $SU(2)$-Higgs model, as the dimensionally reduced
effective theory for QCD \cite{nadkarni90,kajantie97,hart97}. 
A study of the phase structure of the mixed
fundamental and adjoint $SU(2)$ gauge theory \cite{akeyo93b},
which is the lowest order correction to pure $SU(2)$ gauge theory
from a weak coupling expansion in $\beta_h$ of the adjoint
Higgs model \cite{nadkarni90}, has already been successfully
carried out using similar techniques.

However, given the remarkable level of precision that Monte Carlo techniques
have been able to attain in this area, it is perhaps more appropriate
to pursue the study of problems 
which are rather less accessible to such
simulations. For instance, the possibility exists for studying the full $4D$ 
finite temperature theory directly \cite{tan89}. Furthermore,
study of the $U(1)$-Higgs model, with the inclusion of a chemical potential
in the context of superconductivity, appears feasible whereas, in contrast,
the additional phase introduced by the chemical potential 
renders Monte Carlo simulations for this system very problematic.

\section*{Acknowledgements}
The financial support of T.S.E by the Royal Society, and A.R. by the 
Commonwealth Scholarship Commission and the British Council, is
gratefully acknowledged. This work was supported in part by 
the European Commission under the Human Capital and Mobility
programme, contract number CHRX-CT94-0423.

\appendix
\section{Calculational Techniques and Expectation Values}

In this appendix we briefly discuss certain aspects of the  
techniques used for evaluation of the expectation values within the LDE.
All the required expectations may be reduced to the form
\be
 <X>_0 & = & \frac{1}{Z_0}\int[dU][d\rh]X e^{S_0},
\ee
where in all cases $X$
corresponds to a product of real constants, factors of $\rh$, and
$SU(2)$ characters $\ch_r(U)=({\rm Tr} U)_r$. In addition $S_0$, up 
to terms in $\rh$, also consists of fundamental $SU(2)$ characters. 
Thus the exponential may be represented as a character expansion
\be
 e^{\frac{1}{2}J\ch_{1/2}(U)} & = & 
        \frac{2}{J}\sum_{r=0}I_{r+1}(J)\ch_{r/2}(U),
\ee
where $I_r(J)$ are modified Bessel fuctions. 

Almost all the required expectation values $X$ may be converted 
into simple character factors for each link using a combination 
of Clebsch-Gordan expansions, and the following relations, which 
follow easily using the invariance
of the Haar measure and $S_0$ under a unitary 
transformation \cite{zheng87}:
\be
 <\ch_{r/2}(U^{2^n})> & = & (r+1)^{1-2^n} <\ch_{r/2}(U)>^{2^n}. \label{cut}
\ee
Note that these relations also hold when there are additional single--link
characters of $U$ present in the expectation. 
With these techniques most of the
required expectations are reduced to a product of constants
and simple character factors. The group integration then becomes
trivial via use of the orthogonality relation
\be
 \int[dU]\ch_r(UV)\ch_{r'}(U^{-1}W) & = & 
       \frac{1}{{\rm dim}(r)}\de^{rr'}\ch_r(VW).
\ee
The remaining integration over $\rh$ may then be evaluated numerically
using (\ref{Anr}). As an example, consider evaluation of the following
expectation value:
\be
 \left<D_4\right>_0 & = & \frac{1}{4}
      \left<\rh_1^2\ch_{1/2}(U_1U_2U_3^{\dagger}U_4^{\dagger})
               \ch_{1/2}(U_1)\right>_0 \\
    & = & \frac{1}{2^5} \left<\rh_1^2\ch_{1/2}(U_2)\ch_{1/2}(U_3^{\dagger})
         \ch_{1/2}(U_4^{\dagger})
               (1+\ch_1(U_1))\right>_0 \\
    & = & \frac{1}{2^5} (B^0_1)^3(B^1_0+B^1_2).
\ee
In the second line we have made use of (\ref{cut}), and a Clebsch-Gordan
expansion for $(\ch_{1/2}(U))^2$.

However, there are certain diagrams appearing at third order in $\de$
which cannot be reduced to simple character factors for each link due
to the degree of connectedness of the plaquettes. For example,
see diagrams d20, d35 and d36. In order to calculate the required expectations
in this case we use the projection operator technique elaborated
in the Appendix of \cite{akeyo95}, which allows the direct evaluation
of un-traced products of a single link. We refer the reader to \cite{akeyo95}
for more details.

For each diagram required for the analysis in Section 4 the disconnected 
expectation values ($\left<D_i\right>$) calculated in the manner
described above, and
also the corresponding multiplicities ($m_i$), are now presented in the 
following table.
For convenience we define the constants $d_1=2d-3$, $d_2=2d-4$, $d_3=d-1$, 
and $d_4=2d-1$.

\pagebreak

\begin{picture}(400,500)
\thicklines

\newsavebox{\plaqs}
\newsavebox{\plaqtwo}
\newsavebox{\plaqthree}
\newsavebox{\plaqsl}
\newsavebox{\plaqbox}
\newsavebox{\dumbs}
\newsavebox{\dumbtwo}
\newsavebox{\links}

\savebox{\plaqs}{\begin{picture}(15,15)
\put(0,0){\framebox(15,15){}}
\end{picture}}

\savebox{\plaqtwo}{\begin{picture}(15,15)
\put(0,0){\framebox(15,15){2}}
\end{picture}}

\savebox{\plaqthree}{\begin{picture}(15,15)
\put(0,0){\framebox(15,15){3}}
\end{picture}}

\savebox{\plaqsl}{\begin{picture}(25,25)
\put(0,9){\line(4,-3){24}}
\put(12,0){\line(0,2){12}}
\put(0,9){\line(0,2){12}}
\put(0,21){\line(4,-3){24}}
\put(24,-9){\line(0,2){12}}
\put(12,0){\line(3,4){9}}
\put(12,12){\line(3,4){9}}
\put(21,12){\line(0,2){12}}
\end{picture}}

\savebox{\plaqbox}{\begin{picture}(25,25)
\put(0,9){\line(4,-3){12}}
\put(12,0){\line(0,2){12}}
\put(0,9){\line(0,2){12}}
\put(0,21){\line(4,-3){12}}
\put(12,0){\line(3,4){9}}
\put(12,12){\line(3,4){9}}
\put(0,21){\line(3,4){9}}
\put(21,12){\line(0,2){12}}
\put(9,33){\line(4,-3){12}}
\end{picture}}

\savebox{\dumbs}{\begin{picture}(4,15)
\put(2,0){\line(0,2){15}}
\put(2,0){\circle*{4}}
\put(2,15){\circle*{4}}
\end{picture}}

\savebox{\links}{\begin{picture}(4,15)
\put(2,2){\line(0,2){11}}
\end{picture}}
 

\put(-40,500){\line(1,0){480}}
\put(-40,497){\line(1,0){480}}
\put(-35,485){$d\#$}
\put(40,485){$D_i$}
\put(130,485){$m_i$}
\put(320,485){$\left<D_i\right>_0$}
\put(-40,480){\line(1,0){480}}
\put(-40,478){\line(1,0){480}}

 
\put(-35,460){d1}
\put(40,455){\usebox{\dumbs}}
\put(130,460){1}
\put(320,460){$\frac{1}{2}B_1^1$}

\put(-35,430){d2}
\put(34,425){\usebox{\plaqs}}
\put(130,430){$1$}
\put(320,430){$\frac{1}{16}(B^0_1)^4$}

\put(-35,400){d3}
\put(40,395){\usebox{\links}}
\put(130,400){$1$}
\put(320,400){$\frac{1}{2}B^0_1$}
 
\put(-35,370){d4}
\put(30,365){\usebox{\dumbs}}
\put(36,365){\usebox{\plaqs}}
\put(130,370){$4$}
\put(290,370){$\frac{1}{32}(B_0^1)^3(B^1_0+B^1_2)$}

\put(-35,340){d5}
\put(38,335){\usebox{\dumbs}}
\put(42,335){\usebox{\links}}
\put(130,340){$1$}
\put(305,340){$\frac{1}{4}(B^1_0+B^1_2)$}

\put(-35,310){d6}
\put(38,305){\usebox{\plaqtwo}}
\put(130,310){$1$}
\put(300,310){$\frac{1}{4}(1+\frac{1}{27}(B^0_2)^4)$}

\put(-35,280){d7}
\put(30,275){\usebox{\plaqs}}
\put(45,275){\usebox{\plaqs}}
\put(130,280){$4d_1$}
\put(305,280){$\frac{1}{256}(B^0_1)^6(1+B^0_2)$}

\put(-35,250){d8}
\put(38,245){\usebox{\plaqs}}
\put(55,245){\usebox{\links}}
\put(130,250){$4$}
\put(305,250){$\frac{1}{4}(1+\frac{1}{27}(B^0_2)^4)$}

\put(-35,220){d9}
\put(45,215){\usebox{\links}}
\put(50,215){\usebox{\links}}
\put(130,220){$1$}
\put(305,220){$\frac{1}{4}(1+B^0_2))$}

\put(-35,190){d10}
\put(30,185){\usebox{\dumbs}}
\put(36,185){\usebox{\plaqtwo}}
\put(130,190){$4$}
\put(280,190){$\frac{1}{8}B_1^1+\frac{1}{216}(B^0_2)^3(B^1_1+B^1_3)$}

\put(-35,160){d11}
\put(20,155){\usebox{\dumbs}}
\put(26,155){\usebox{\plaqs}}
\put(41,155){\usebox{\plaqs}}
\put(120,160){$24d_1$}
\put(280,160){$\frac{1}{512}(B_0^1)^5(1+B^0_2)(B^1_0+B^1_2)$}

\put(-35,130){d12}
\put(39,125){\usebox{\dumbs}}
\put(26,125){\usebox{\plaqs}}
\put(41,125){\usebox{\plaqs}}
\put(125,130){$4d_1$}
\put(290,130){$\frac{1}{512}(B_0^1)^6(2B^1_1+B^1_3)$}

\put(-35,100){d13}
\put(30,95){\usebox{\plaqs}}
\put(46,95){\usebox{\links}}
\put(50,95){\usebox{\dumbs}}
\put(130,100){$4$}
\put(290,100){$\frac{1}{64}(B_0^1)^3(2B^1_1+B^1_3)$}

\put(-35,70){d14}
\put(28,65){\usebox{\dumbs}}
\put(34,65){\usebox{\plaqs}}
\put(50,65){\usebox{\links}}
\put(130,70){$12$}
\put(280,70){$\frac{1}{64}(B_0^1)^2(1+B^0_2)(B^1_0+B^1_2)$}

\put(-35,40){d15}
\put(35,35){\usebox{\dumbs}}
\put(40,35){\usebox{\links}}
\put(45,35){\usebox{\links}}
\put(130,40){$1$}
\put(300,40){$\frac{1}{8}(2B^1_1+B^1_3)$}

\put(-35,10){d16}
\put(38,5){\usebox{\plaqthree}}
\put(130,10){$1$}
\put(285,10){$\frac{1}{32}((B^0_1)^4+\frac{1}{16}(B^0_3)^4)$}

\put(-35,-20){d17}
\put(30,-25){\usebox{\plaqtwo}}
\put(45,-25){\usebox{\plaqs}}
\put(120,-20){$12d_1$}
\put(265,-20){$\frac{1}{2^6}(1+\frac{1}{3^3}(B_0^1)^3(B^0_2)^3(B^0_1+B^0_3))$}

\put(-35,-50){d18}
\put(20,-55){\usebox{\plaqs}}
\put(35,-55){\usebox{\plaqs}}
\put(50,-55){\usebox{\plaqs}}
\put(100,-50){$36d_1^2-24d_2$}
\put(290,-50){$\frac{1}{2^{12}}(B^0_1)^8(1+B^0_2)^2$}

\put(-35,-90){d19}
\put(30,-95){\usebox{\plaqsl}}
\put(115,-90){$4d_1d_2$}
\put(290,-90){$\frac{1}{2^{12}}(B^0_1)^9(2B^0_1+B^0_3)$}

\put(-35,-130){d20}
\put(30,-145){\usebox{\plaqbox}}
\put(120,-130){$8d_2$}
\put(270,-130){$\frac{1}{3^22^{10}}(B^0_1)^6(2(B^0_2)^3+9(B^0_2)^2+9)$}

\put(-40,-150){\line(0,2){650}}
\put(440,-150){\line(0,2){650}}
\put(-40,-150){\line(2,0){480}}

\end{picture}

\pagebreak

\begin{picture}(400,500)
\thicklines

\savebox{\plaqs}{\begin{picture}(15,15)
\put(0,0){\framebox(15,15){}}
\end{picture}}

\savebox{\plaqtwo}{\begin{picture}(15,15)
\put(0,0){\framebox(15,15){2}}
\end{picture}}

\savebox{\plaqthree}{\begin{picture}(15,15)
\put(0,0){\framebox(15,15){3}}
\end{picture}}

\savebox{\plaqsl}{\begin{picture}(25,25)
\put(0,9){\line(4,-3){24}}
\put(12,0){\line(0,2){12}}
\put(0,9){\line(0,2){12}}
\put(0,21){\line(4,-3){24}}
\put(24,-9){\line(0,2){12}}
\put(12,0){\line(3,4){9}}
\put(12,12){\line(3,4){9}}
\put(21,12){\line(0,2){12}}
\end{picture}}

\savebox{\plaqbox}{\begin{picture}(25,25)
\put(0,9){\line(4,-3){12}}
\put(12,0){\line(0,2){12}}
\put(0,9){\line(0,2){12}}
\put(0,21){\line(4,-3){12}}
\put(12,0){\line(3,4){9}}
\put(12,12){\line(3,4){9}}
\put(0,21){\line(3,4){9}}
\put(21,12){\line(0,2){12}}
\put(9,33){\line(4,-3){12}}
\end{picture}}

\savebox{\dumbs}{\begin{picture}(4,15)
\put(2,0){\line(0,2){15}}
\put(2,0){\circle*{4}}
\put(2,15){\circle*{4}}
\end{picture}}

\savebox{\links}{\begin{picture}(4,15)
\put(2,2){\line(0,2){11}}
\end{picture}}
 

\put(-40,500){\line(1,0){480}}
\put(-40,497){\line(1,0){480}}
\put(-35,485){$d\#$}
\put(40,485){$D_i$}
\put(130,485){$m_i$}
\put(320,485){$\left<D_i\right>_0$}
\put(-40,480){\line(1,0){480}}
\put(-40,478){\line(1,0){480}}

 
\put(-35,460){d21}
\put(30,455){\usebox{\plaqtwo}}
\put(48,455){\usebox{\links}}
\put(115,460){4}
\put(250,460){$\frac{1}{8}(B^0_1+\frac{1}{3^3}(B^0_2)^3(B^0_1+B^0_3))$}

\put(-35,430){d22}
\put(25,425){\usebox{\plaqs}}
\put(40,425){\usebox{\plaqs}}
\put(58,425){\usebox{\links}}
\put(110,430){$24d_1$}
\put(270,430){$\frac{1}{2^9}(B^0_1)^5(1+B^0_2)^2$}

\put(-35,400){d23}
\put(30,395){\usebox{\plaqs}}
\put(45,395){\usebox{\plaqs}}
\put(47,395){\usebox{\links}}
\put(110,400){$4d_1$}
\put(270,400){$\frac{1}{2^9}(B^0_1)^6(2B^0_1+B^0_3)$}

\put(-35,370){d24}
\put(30,365){\usebox{\plaqs}}
\put(48,365){\usebox{\links}}
\put(51,365){\usebox{\links}}
\put(115,370){$4$}
\put(270,370){$\frac{1}{2^6}(B^0_1)^3(2B^0_1+B^0_3)$}

\put(-35,340){d25}
\put(36,335){\usebox{\plaqs}}
\put(30,335){\usebox{\links}}
\put(53,335){\usebox{\links}}
\put(115,340){$12$}
\put(270,340){$\frac{1}{2^6}(B^0_1)^2(1+B^0_2)^2$}

\put(-35,310){d26}
\put(40,305){\usebox{\links}}
\put(44,305){\usebox{\links}}
\put(48,305){\usebox{\links}}
\put(115,310){$1$}
\put(270,310){$\frac{1}{8}(2B^0_1+B^0_3)$}

\put(-35,280){d27}
\put(30,275){\usebox{\plaqthree}}
\put(48,275){\usebox{\dumbs}}
\put(115,280){$4$}
\put(210,280){$\frac{1}{64}(B^0_1)^3(B^1_0+B^1_2)
        +\frac{1}{2^{10}}(B^0_3)^3(B^1_2+B^1_4)$}

\put(-35,250){d28}
\put(25,245){\usebox{\plaqtwo}}
\put(40,245){\usebox{\plaqs}}
\put(58,245){\usebox{\dumbs}}
\put(110,250){$36d_1$}
\put(205,250){$\frac{1}{2^7}(B^0_1)^2(B^1_0+B^1_2)
        (B^0_1+\frac{1}{3^3}(B^0_2)^3(B^0_1+B^0_3))$}

\put(-35,220){d29}
\put(25,215){\usebox{\plaqs}}
\put(40,215){\usebox{\plaqtwo}}
\put(58,215){\usebox{\dumbs}}
\put(110,220){$36d_1$}
\put(200,220){$\frac{1}{2^7}(B^0_1)^3(B^0_1B^1_1+\frac{1}{3^3}
        (B^0_2)^2(B^0_1+B^0_3)(B^1_1+B^1_3))$}

\put(-35,190){d30}
\put(30,185){\usebox{\plaqs}}
\put(45,185){\usebox{\plaqtwo}}
\put(43,185){\usebox{\dumbs}}
\put(110,190){$12d_1$}
\put(205,190){$\frac{1}{2^7}(B^0_1)^3(B^1_0+B^1_2+\frac{1}{3^3}
        (B^0_2)^3(B^1_0+2B^1_2+B^1_4))$}

\put(-35,160){d31}
\put(20,155){\usebox{\plaqs}}
\put(35,155){\usebox{\plaqs}}
\put(50,155){\usebox{\plaqs}}
\put(68,155){\usebox{\dumbs}}
\put(95,160){$288d_1^2-192d_2$}
\put(250,160){$\frac{1}{2^{13}}(B^0_1)^7(1+B^0_2)^2(B^1_0+B^1_2)$}

\put(-35,130){d32}
\put(20,125){\usebox{\plaqs}}
\put(35,125){\usebox{\plaqs}}
\put(50,125){\usebox{\plaqs}}
\put(48,125){\usebox{\dumbs}}
\put(100,130){$72d_1^2-48d_2$}
\put(250,130){$\frac{1}{2^{13}}(B^0_1)^8(1+B^0_2)(2B^1_1+B^1_3)$}

\put(-35,90){d33}
\put(30,85){\usebox{\plaqsl}}
\put(25,94){\usebox{\dumbs}}
\put(110,90){$36d_1d_2$}
\put(245,90){$\frac{1}{2^{13}}(B^0_1)^8(B^1_0+B^1_2)(2B^0_1+B^0_3)$}

\put(-35,50){d34}
\put(30,45){\usebox{\plaqsl}}
\put(37,43){\usebox{\dumbs}}
\put(110,50){$4d_1d_2$}
\put(250,50){$\frac{1}{2^{13}}(B^0_1)^9(2B^1_0+3B^1_2+B^1_4)$}

\put(-35,10){d35}
\put(30,0){\usebox{\plaqbox}}
\put(25,8){\usebox{\dumbs}}
\put(110,10){$48d_2$}
\put(205,10){$\frac{1}{3^22^{10}}(B^0_1)^5(B^1_0+B^1_2)
            (2(B^0_2)^3+9(B^0_2)^2+9)$}

\put(-35,-30){d36}
\put(30,-40){\usebox{\plaqbox}}
\put(40,-40){\usebox{\dumbs}}
\put(110,-30){$24d_2$}
\put(165,-30){$\frac{1}{3^22^{12}}(B^0_1)^6(4B^0_2B^1_3(B^0_2+3)
         +2B^1_1(5(B^0_2)^2+6B^0_2+9))$}

\put(-35,-60){d37}
\put(30,-65){\usebox{\plaqtwo}}
\put(48,-65){\usebox{\links}}
\put(51,-65){\usebox{\dumbs}}
\put(115,-60){$4$}
\put(210,-60){$\frac{1}{16}(B^1_0+B^1_2+\frac{1}{3^3}(B^0_2)^3
        (B^1_0+2B^1_2+B^1_4))$}

\put(-35,-90){d38}
\put(30,-95){\usebox{\plaqtwo}}
\put(25,-95){\usebox{\links}}
\put(48,-95){\usebox{\dumbs}}
\put(115,-90){$12$}
\put(210,-90){$\frac{1}{16}(B^1_0B^0_1+\frac{1}{3^3}(B^0_2)^2
       (B^0_1+B^0_3)(B^1_1+B^1_3))$}

\put(-35,-120){d39}
\put(25,-125){\usebox{\plaqs}}
\put(40,-125){\usebox{\plaqs}}
\put(58,-125){\usebox{\links}}
\put(61,-125){\usebox{\dumbs}}
\put(115,-120){$24d_1$}
\put(245,-120){$\frac{1}{2^{10}}(B^0_1)^5(1+B^0_2)(2B^1_1+B^1_3)$}

\put(-40,-140){\line(0,2){640}}
\put(440,-140){\line(0,2){640}}
\put(-40,-140){\line(2,0){480}}

\end{picture}

\pagebreak

\begin{picture}(400,500)
\thicklines

\savebox{\plaqs}{\begin{picture}(15,15)
\put(0,0){\framebox(15,15){}}
\end{picture}}

\savebox{\plaqtwo}{\begin{picture}(15,15)
\put(0,0){\framebox(15,15){2}}
\end{picture}}

\savebox{\plaqthree}{\begin{picture}(15,15)
\put(0,0){\framebox(15,15){3}}
\end{picture}}

\savebox{\plaqsl}{\begin{picture}(25,25)
\put(0,9){\line(4,-3){24}}
\put(12,0){\line(0,2){12}}
\put(0,9){\line(0,2){12}}
\put(0,21){\line(4,-3){24}}
\put(24,-9){\line(0,2){12}}
\put(12,0){\line(3,4){9}}
\put(12,12){\line(3,4){9}}
\put(21,12){\line(0,2){12}}
\end{picture}}

\savebox{\plaqbox}{\begin{picture}(25,25)
\put(0,9){\line(4,-3){12}}
\put(12,0){\line(0,2){12}}
\put(0,9){\line(0,2){12}}
\put(0,21){\line(4,-3){12}}
\put(12,0){\line(3,4){9}}
\put(12,12){\line(3,4){9}}
\put(0,21){\line(3,4){9}}
\put(21,12){\line(0,2){12}}
\put(9,33){\line(4,-3){12}}
\end{picture}}

\savebox{\dumbs}{\begin{picture}(4,15)
\put(2,0){\line(0,2){15}}
\put(2,0){\circle*{4}}
\put(2,15){\circle*{4}}
\end{picture}}

\savebox{\dumbtwo}{\begin{picture}(15,15)
\put(0,0){\line(0,2){15}}
\put(0,15){\line(2,0){15}}
\put(0,0){\circle*{4}}
\put(0,15){\circle*{4}}
\put(15,15){\circle*{4}}
\end{picture}}

\savebox{\links}{\begin{picture}(4,15)
\put(2,2){\line(0,2){11}}
\end{picture}}
 

\put(-40,500){\line(1,0){480}}
\put(-40,497){\line(1,0){480}}
\put(-35,485){$d\#$}
\put(40,485){$D_i$}
\put(130,485){$m_i$}
\put(320,485){$\left<D_i\right>_0$}
\put(-40,480){\line(1,0){480}}
\put(-40,478){\line(1,0){480}}

 
\put(-35,460){d40}
\put(25,455){\usebox{\plaqs}}
\put(40,455){\usebox{\plaqs}}
\put(58,455){\usebox{\dumbs}}
\put(42,455){\usebox{\links}}
\put(120,460){$24d_1$}
\put(250,460){$\frac{1}{2^{10}}(B^0_1)^5(2B^0_1+B^0_3)(B^1_0+B^1_2)$}

\put(-35,430){d41}
\put(25,425){\usebox{\plaqs}}
\put(40,425){\usebox{\plaqs}}
\put(38,425){\usebox{\dumbs}}
\put(58,425){\usebox{\links}}
\put(120,430){$24d_1$}
\put(250,430){$\frac{1}{2^{10}}(B^0_1)^5(2B^1_1+B^1_3)(1+B^0_2)$}

\put(-35,400){d42}
\put(25,395){\usebox{\plaqs}}
\put(40,395){\usebox{\plaqs}}
\put(38,395){\usebox{\dumbs}}
\put(43,395){\usebox{\links}}
\put(120,400){$4d_1$}
\put(255,400){$\frac{1}{2^{10}}(B^0_1)^6(2B^1_0+3B^1_2+B^1_4)$}

\put(-35,370){d43}
\put(25,365){\usebox{\plaqs}}
\put(40,365){\usebox{\plaqs}}
\put(58,365){\usebox{\dumbs}}
\put(20,365){\usebox{\links}}
\put(120,370){$120d_1$}
\put(250,370){$\frac{1}{2^{10}}(B^0_1)^4(1+B^0_2)^2(B^1_0+B^1_2)$}

\put(-35,340){d44}
\put(30,335){\usebox{\plaqs}}
\put(48,335){\usebox{\links}}
\put(58,335){\usebox{\dumbs}}
\put(53,335){\usebox{\links}}
\put(130,340){$4$}
\put(260,340){$\frac{1}{2^7}(B^0_1)^3(2B^1_0+3B^1_2+B^1_4)$}

\put(-35,310){d45}
\put(35,305){\usebox{\plaqs}}
\put(53,305){\usebox{\links}}
\put(58,305){\usebox{\dumbs}}
\put(30,305){\usebox{\links}}
\put(130,310){$24$}
\put(260,310){$\frac{1}{2^7}(B^0_1)^2(2B^1_1+B^1_3)(1+B^0_2)$}

\put(-35,275){d46}
\put(35,270){\usebox{\plaqs}}
\put(37,288){\line(2,0){11}}
\put(53,270){\usebox{\dumbs}}
\put(30,270){\usebox{\links}}
\put(130,275){$24$}
\put(270,275){$\frac{1}{2^7}B^0_1(2B^1_0+B^1_2)(1+B^0_2)^2$}

\put(-35,245){d47}
\put(35,240){\usebox{\links}}
\put(40,240){\usebox{\links}}
\put(50,240){\usebox{\dumbs}}
\put(45,240){\usebox{\links}}
\put(130,245){$1$}
\put(280,245){$\frac{1}{16}(2B^1_0+3B^1_2+B^1_4)$}

\put(-35,215){d48}
\put(40,210){\usebox{\dumbs}}
\put(45,210){\usebox{\dumbs}}
\put(130,215){$1$}
\put(300,215){$\frac{1}{4}(B^2_0+B^2_2)$}

\put(-35,185){d49}
\put(30,180){\usebox{\dumbs}}
\put(35,180){\usebox{\dumbs}}
\put(40,180){\usebox{\plaqs}}
\put(130,185){$4$}
\put(275,185){$\frac{1}{2^6}(B^0_1)^3(2B^2_1+B^2_3)$}

\put(-35,155){d50}
\put(40,150){\usebox{\dumbtwo}}
\put(40,150){\usebox{\plaqs}}
\put(130,155){$8$}
\put(280,155){$\frac{1}{2^6}(B^0_1)^2(B^1_0+B^1_2)^2$}

\put(-35,125){d51}
\put(45,120){\usebox{\dumbtwo}}
\put(30,120){\usebox{\plaqs}}
\put(120,125){$32d_3$}
\put(270,125){$\frac{1}{2^6}(B^0_1)^3B^1_1(B^1_0+B^1_2)$}

\put(-35,95){d52}
\put(40,90){\usebox{\dumbs}}
\put(45,90){\usebox{\dumbs}}
\put(50,90){\usebox{\links}}
\put(130,95){$1$}
\put(290,95){$\frac{1}{8}(2B^2_1+B^2_3)$}

\put(-35,65){d53}
\put(40,60){\usebox{\links}}
\put(45,60){\usebox{\dumbtwo}}
\put(120,65){$4d_4$}
\put(290,65){$\frac{1}{8}B^1_1(B^1_0+B^1_2)$}

\put(-40,30){\line(0,2){470}}
\put(440,30){\line(0,2){470}}
\put(-40,30){\line(2,0){480}}

\end{picture}

\pagebreak

\section{Expansion of $A^n_r$, and Higher Order Corrections}

After the group integrations have been performed, all expectation values
up to $O(\de^2)$ reduce to a series of ratios of the integral (\ref{Anr}),
\be
 (A^n_r)^d & = & \int_0^{\infty}{\mbox d}\rh \rh^{2n+3}e^{-V_H}
              \left((r+1)\frac{I_{r+1}(J+\beta_h\rh^2)}{\frac{1}{2}
               (J+\beta_h\rh^2)}\right)^d, \label{Anrd}
\ee
for various choices of $n$ and $r$ depending on the diagram being evaluated.
While the integral may easily be performed numerically, it is 
useful to consider an analytic series solution, particularly in the
strong coupling regime $\beta\rightarrow 0$, and in the regime of a minimal
gauge-Higgs coupling, $\beta_h\ll 1$. 

As was discussed in Section 4, the LDE approach naturally includes a
solution at $\tilde J=0$ which corresponds to the strong coupling
expansion appropriate for $\beta\ll 1$. We now wish to express
(\ref{Anrd}) as an expansion in $\beta_h$. Expanding the modified Bessel
function in a power series, we may represent the integral in the form
\be
 (A^n_r)^d & = & \int_0^{\infty}{\mbox d}\rh \rh^{2n+3}e^{-V_H}
        \left[\sum_{k=0}^{\infty}\la_k\right]^d \\
   & = & \int_0^{\infty}{\mbox d}\rh \rh^{2n+3}e^{-V_H}
        \sum_{m=0}^{\infty}c_m \label{raise}
\ee
where
\be
 \la_k & = & \frac{r+1}{k!(k+r+1)!} 
    \left(\frac{J+\beta_h\rh^2}{2}\right)^{r+2k}, \label{lak}
\ee
and in (\ref{raise}) we have expressed the $d^{th}$ power of the series in 
$\la_k$ as a new series with terms $c_m$ which satisfy the 
recursion relation,
\be
 c_m & = & \frac{1}{m\la_0}\sum_{p=1}^m (pd-m+p)\la_p c_{m-p},
\ee
with initial condition $c_0=\la_0^d$. This recursion
relation may be solved by noting that the structure implies
that $c_m$ is given by a sum of products of the $\la_k$ whose
exponents in each monomial are given by an appropriate partition
of $m$. The structure of the coefficients of these monomials is then
easily deduced, and one finds that the terms $c_m$ have the general
form
\be
 c_m & = & \sum_{\{i_p\}}^{\sum_{p=1}^m p i_p=m}
             \left.\frac{d!}{\prod_{p=1}^m i_p!
         \left(d-\sum_{p=1}^m i_p\right)!}\la_0^{d-\sum_{p=1}^m i_p}
           \prod_{p=1}^m \la_p^{i_p}\right.. \label{cm}
\ee
In the strong coupling limit with $J=0$ one may then readily perform
the integration over $\rh$ to obtain $A^n_r$ as a power series in $\beta_h$.
In the general case, after forming the product of the factors $\la_p^{i_p}$
in (\ref{cm}),
we expand $\la_k$ in (\ref{lak}) as a power series in $\beta_h$, giving
$c_m$ as an explicit power series
in $\beta_h$. For compactness in the following results we
define $\si_t\equiv \sum_{q=1}^m(q)^ti_q$, where the 
$i_q$ are determined by the relevant partition of $m$. With this definition,
the result for $c_m$ becomes
\be
 c_m & = & \sum_{\{i_p|\si_1=m\}} \left.P(d,m,r)
     \left(\frac{J}{2}\right)^{rd+2\si_1}\sum_{s=0}^{rd+2\si_1}
      \left(\frac{rd+2\si_1}{s}\right)\left(\frac{\beta_h}{J}\right)^s\rh^{2s}
           \right.,
\ee
where
\be
 P(d,m,r) & \equiv & \frac{d!(r+1)^{\si_0}}{(r!)^{d-\si_0}(d-\si_0)! 
          \prod_{t=1}^m i_t!(t!(t+r+1)!)^{i_t}}.
\ee
Returning to the full integral, the $\rh$ integration may now be performed
using the relation 
\be
 \int_0^{\infty}{\mbox d}\rh \rh^{2x+3}e^{-V_H(\beta_r)}
     & = & \frac{1}{2}(x+1)!C(\beta_r,2x),
\ee
where $C$ may be written in terms of a parabolic cylinder function
$D_n(x)$ as 
\be
 C(\beta_r,x) & = & e^{\frac{(1-2\beta_r)^2}{8\beta_r}-\beta_r}
    (2\beta_r)^{-1-\frac{x}{4}} D_{-\frac{x}{2}-2}\left(
         \frac{1-2\beta_r}{\sqrt{2\beta_r}}\right),
\ee
which tends to unity as $\beta_r\rightarrow 0$. Finally, the full integral
is  given by the power series
\be
 (A^n_r)^d & = & \sum_{m=0}^{\infty} \sum_{\{i_p|\si_1=m\}}
        P(d,m,r)\left(\frac{J}{2}\right)^{rd+2\si_1} \nonumber\\
   & & \;\;\;\;\;\;\;\;\;
      \times\frac{1}{2}\sum_{s=0}^{rd+2\si_1}\left(\frac{rd+2\si_1}{s}\right)
     \left(\frac{\beta_h}{J}\right)^s Q(\beta_r,2n+2s),
\ee
where $Q(\beta_r,2x)=(x+1)!C(\beta_r,2x)/2$.
In the strong coupling limit ($J=0$) this reduces to 
\be
 (A^n_r)_S^d & = & \sum_{m=0}^{\infty}\left. \sum_{\{i_p|\si_1=m\}}
        P(d,m,r)\left(\frac{\beta_h}{2}\right)^{rd+2\si_1}
        Q(\beta_r,2n+rd+2\si_1)\right.
\ee
The lowest order terms of this expansion are given by
\be
 (A^n_r)_S^d & = & \left(\frac{\beta_h}{2}\right)^{rd}\frac{(n+rd/2+1)!}
    {2(r!)^d}\left[1+\left(\frac{\beta_h}{2}\right)^2d\frac{r(r+1)}{(r+2)!}
       (n+rd/2+2)+O\left(\frac{\beta_h}{2}\right)^4\right].
\ee
Thus, if we focus attention on the gauge sector and the lowest order 
contribution to the average plaquette energy, given by $d2$, we find
that in the strong coupling region the $\beta_h$ dependence
is given by
\be
 B^0_1 & = & \frac{\beta_h}{2}\left(\frac{3}{4}\sqrt{\pi}\right)^{1/3}
       \left[1+\frac{3}{2}\left(\frac{\beta_h}{2}\right)^2
          +O\left(\frac{\beta_h}{2}\right)^4\right].
\ee
Therefore, according to the expectation value $\left<D_2\right>$, in this regime 
the leading dependence on $\beta_h$ has the form $(\beta_h/2)^4$, which
is a small correction throughout the symmetric
phase where $\beta_h<0.34$, a result which is borne out by the numerical
analysis presented in Fig.~1. 

While the previous results are exact for expectation values up to $O(\de^2)$,
our restriction of the Higgs modulus in the interaction to the
form $\rh_i\rh_j=\rh_i^2+\de^2\rh_i\De\rh_i$ implies that higher
order calculations of observables will require additional
correction terms for the lower order diagrams. As we have worked 
only to $O(\de^3)$, it is only necessary to consider
the first--order correction to the $O(\de^0)$ and $O(\de)$ contributions
to $\left<E_L\right>$.

Calculation of the correction is conveniently illustrated in the
case of $Z_0$. Expanding the exponential correction term we obtain,
\be
 Z_0 & = & \int[\tilde{{\rm d}\rh}]\prod_{l_{ij}}\int[{\rm d}U_{ij}]
   e^{\frac{1}{2}(J+\beta_h\rh_i^2)\ch_{1/2}(U_{ij})}
   \sum_{n=0}^{\infty}\frac{\de^{2n}}{2^nn!}(\rh_i\rh_j-\rh_i^2)^n
          (\ch_{1/2}(U_{ij}))^n,
\ee
where we define $[\tilde{{\rm d}\rh}]=\prod_i{\rm d}\rh_i \rh_i^3 \exp(-V_H)$.
On performing a character expansion of the exponential, 
$Z_0$ may be expressed in the form
\be
 Z_0 & = & \int[\tilde{{\rm d}\rh}]\prod_{l_{ij}}\sum_{n=0}^{\infty}
    \sum_{r\in{\rm {\bf Z}}_{+}}(r+1)\frac{I_{r+1}(J+\beta_h\rh_i^2)}
       {\frac{1}{2}(J+\beta_h\rh_i^2)}\frac{\de^{2n}}{2^nn!}
    (\rh_i\rh_j-\rh_i^2)^n \int[{\rm d}U_{ij}]
      \ch_{r/2}\sum_{j=0,1/2}^{n/2}c_j\ch_j.
\ee
The final sum begins with $j=0$ ($j=1/2$) for even (odd) $n$, and the
coefficients $c_j$ are determined by the expansion of the
product $(\ch_{1/2}(U_{ij}))^n$. This result is quite general and includes
all higher order corrections. As mentioned earlier, we only
require the first--order correction in the present case, corresponding
to $n=1$. To this order we have
\be
 Z_0 & = & \int[\tilde{{\rm d}\rh}]\prod_i\left(\frac{I_1(J+\beta_h\rh_i^2)}
     {\frac{1}{2}(J+\beta_h\rh_i^2)}\right)^d\prod_{l_{ij}}
      \left[1+\de^2(\rh_i\rh_j-\rh_i^2)\frac{I_2(J+\beta_h\rh_i^2)}
     {I_1(J+\beta_h\rh_i^2)}\right]+O(\de^4).
       \label{del2}
\ee
If we focus on the expression in square brackets, then since the
ratio of Bessel functions has an upper bound of unity for large
values of its argument, one observes that the term of $O(\de^2)$ is indeed
a small correction provided that $\rh_i\sim \rh_j$ for $i$ and $j$ nearest 
neighbour sites. If one makes the assumption that $\rh_i$ and
$\rh_j$ differ by at most 10\%, which seems reasonable for expectation
values dominated by long wavelength modes, then one finds that
the correction is almost negligible despite appearing to be of $O(10\%)$
in (\ref{del2}). The reason is that, for non-zero $\beta_r$, the integral
(\ref{del2}) is dominated by contributions from the region 
$\rh<10-15$ for which
the ratio $I_2/I_1$  is small. Furthermore, the correction needs to be
compared with the disconnected expectation value at the appropriate
order in $\de$ rather than directly with its zeroth order contribution.
Once one does this the correction is indeed found to be small
having negligible effect on the results, as was discussed in Section 4.

This result may be made more explicit by
extracting terms up to $O(\de^2)$ from the product, which
take the form $1+\de^2\sum_{<ij>}(\rh_i\rh_j-\rh_i^2)I_2(x_i)/I_1(x_i)$,
where $x_i\equiv J+\beta_h\rh_i^2$,
\be
 Z_0 & = & \tilde Z_0^{N_s}\left(1+\frac{\de^2}{\tilde Z_0^2} 
       \sum_{<ij>} \tilde Z_{l_{ij}}
        +O(\de^4)\right).
\ee
The first factor is given by the standard zeroth--order contribution,
\be
 \tilde Z_0 & = & \int[\tilde{{\rm d}\rh_i}]\left(\frac{I_1(J+\beta_h\rh_i^2)}
     {\frac{1}{2}(J+\beta_h\rh_i^2)}\right)^d,
\ee
while the $O(\de^2)$ correction also contains a sum of contributions
of the form
\be
 \tilde Z_{1_{ij}} & = & \int[\tilde{{\rm d}\rh_i}\tilde{{\rm d}\rh_j}]
    (\rh_i\rh_j-\rh_i^2)\prod_{p=i,j}
    \left(\frac{I_1(x_p)}{\frac{1}{2}x_p}\right)^d\frac{I_2(x_i)}{I_1(x_i)}.
\ee
A further reduction in the relative magnitude of the correction
would then follow from possible cancellations of terms in the sum over
the links of the lattice, as the factor $(\rh_i\rh_j-\rh_i^2)$
will vary in sign.

\bibliographystyle{prsty}

\begin{thebibliography}{10}

\bibitem{kajantie96a}
K. Kajantie, M. Laine, K. Rummukainen, and M. Shaposhnikov, Nucl. Phys. B {\bf
  {\bf 458}},  90  (1996).

\bibitem{kajantie96d}
K. Kajantie, M. Laine, K. Rummukainen, and M. Shaposhnikov, Nucl. Phys. B {\bf
  {\bf 493}},  413  (1997).

\bibitem{evertz87}
H.~G. Evertz, J. Jers\'ak, and K. Kanaya, Nucl. Phys. B {\bf {\bf 285}},  229
  (1987).

\bibitem{damgaard87}
P. Damgaard and U.~M. Heller, Nucl. Phys. B {\bf {\bf 294}},  253  (1987).

\bibitem{damgaard88}
P. Damgaard and U.~M. Heller, Nucl. Phys. B {\bf {\bf 304}},  63  (1988).

\bibitem{bock90}
W. Bock {\it et~al.}, Phys. Rev. D {\bf {\bf 41}},  2573  (1990).

\bibitem{csikor96}
F. Csikor {\it et~al.}, Nucl. Phys. B {\bf 474},  421  (1996).

\bibitem{aoki97}
Y. Aoki, Nucl. Phys. B (Proc. Suppl.) {\bf 53},  609  (1997).

\bibitem{kajantie96c}
K. Kajantie, M. Laine, K. Rummukainen, and M. Shaposhnikov, Phys. Rev. Lett.
  {\bf {\bf 77}},  2887  (1996).

\bibitem{karsch97}
F. Karsch, T. Neuhaus, A. Patk\'os, and J. Rank, Nucl. Phys. B (Proc. Suppl.)
  {\bf 53},  623  (1997).

\bibitem{gurtler97}
M. G\"urtler, E. Ilgenfritz, and A. Schiller, 
 Preprint: hep-lat/9704013  (1997).

\bibitem{fradkin79}
E. Fradkin and S.~H. Shenker, Phys. Rev. D {\bf {\bf 19}},  3682  (1979).

\bibitem{philipsen96}
O. Philipsen, M. Teper, and H. Wittig, Nucl. Phys. B {\bf {\bf 469}},  445
  (1996).

\bibitem{philipsen97}
O. Philipsen, M. Teper, and H. Wittig, Preprint: hep-lat/9709145  (1997).

\bibitem{reuter93}
M. Reuter and C. Wetterich, Nucl. Phys. B {\bf 408},  91  (1993).

\bibitem{tetradis97}
N. Tetradis, Nucl. Phys. B {\bf 488},  92  (1997).

\bibitem{buchmuller94}
W. Buchm\"uller and O. Philipsen, Nucl. Phys. B {\bf 443},  47  (1994).

\bibitem{ejr97a}
T.~S. Evans, H.~F. Jones, and A. Ritz, hep-th/9707539, to appear in:
  Proceedings of Eotvos Conference in Science: Strong and Electroweak Matter
  (SEWM 97), Eger, Hungary, 21-25 May 1997  (1997).

\bibitem{zheng87}
X.-T. Zheng, Z.~G. Tan, and J. Wang, Nucl. Phys. B {\bf {\bf 287}},  171
  (1987).

\bibitem{tan89}
C.-I. Tan and X.-T. Zheng, Phys. Rev. D {\bf {\bf 39}},  623  (1989).

\bibitem{duncan88}
A. Duncan and M. Moshe, Phys. Lett. B {\bf {\bf 215}},  352  (1988).

\bibitem{duncan89}
A. Duncan and H.~F. Jones, Nucl. Phys. B {\bf {\bf 320}},  189  (1989).

\bibitem{buckley92a}
I.~R.~C. Buckley and H.~F. Jones, Phys. Rev. D {\bf {\bf 45}},  654  (1992).

\bibitem{buckley92b}
I.~R.~C. Buckley and H.~F. Jones, Phys. Rev. D {\bf {\bf 45}},  2073  (1992).

\bibitem{akeyo93a}
J.~O. Akeyo and H.~F. Jones, Phys. Rev. D {\bf 47},  1668  (1993).

\bibitem{akeyo93b}
J.~O. Akeyo and H.~F. Jones, Z. Phys. C {\bf 58},  629  (1993).

\bibitem{zheng91}
X.-T. Zheng and B.-S. Liu, Int. J. Mod. Phys. A {\bf {\bf 6}},  103  (1991).

\bibitem{yang91}
J.~M. Yang, J. Phys. G {\bf {\bf 17}},  L143  (1991).

\bibitem{yang92}
J.~M. Yang, C.~M. Wu, and P.~Y. Zhao, J. Phys. G {\bf {\bf 18}},  L1  (1992).

\bibitem{akeyo95}
J.~O. Akeyo, C.~S. Parker, and H.~F. Jones, Phys. Rev. D {\bf 51},  1298
  (1995).

\bibitem{kajantie96b}
K. Kajantie, M. Laine, K. Rummukainen, and M. Shaposhnikov, Nucl. Phys. B {\bf
  {\bf 466}},  189  (1996).

\bibitem{stevenson81}
P.~M. Stevenson, Phys. Rev. D {\bf {\bf 23}},  2916  (1981).

\bibitem{buckley93}
I.~R.~C. Buckley, A. Duncan, and H.~F. Jones, Phys. Rev. D {\bf {\bf 47}},
  2554  (1993).

\bibitem{duncan93}
A. Duncan and H.~F. Jones, Phys. Rev. D {\bf {\bf 47}},  2560  (1993).

\bibitem{frohlich81}
J. Fr\"ohlich, G. Morchio, and F. Strocchi, Nucl. Phys. B {\bf {\bf 190}},  553
   (1981).

\bibitem{karsch96}
F. Karsch, T. Neuhaus, A. Patk\'os, and J. Rank, Nucl. Phys. B {\bf 474},  217
  (1996).

\bibitem{buchmuller97}
W. Buchm\"uller and O. Philipsen, Phys. Lett. B {\bf 397},  112  (1997).

\bibitem{halliday84}
I.~G. Halliday, Rep. Prog. Phys. {\bf {\bf 47}},  987  (1984).

\bibitem{damgaard85}
P. Damgaard and U.~M. Heller, Phys. Lett. B {\bf {\bf 164}},  121  (1985).

\bibitem{gurtler96}
M. G\"urtler {\it et~al.}, Nucl. Phys. B {\bf {\bf 483}},  383  (1997).

\bibitem{nadkarni90}
S. Nadkarni, Nucl. Phys. B {\bf {\bf 334}},  559  (1990).

\bibitem{kajantie97}
K. Kajantie, M. Laine, K. Rummukainen, and M. Shaposhnikov, Preprint:
  hep-ph/9704416  (1997).

\bibitem{hart97}
A. Hart, O. Philipsen, J.~D. Stack, and M. Teper, Phys. Lett. B {\bf {\bf
  396}},  217  (1997).

\end{thebibliography}

\end{document}